\input harvmac
\input epsf
\epsfverbosetrue

\def\p{\partial}
\def\ap{\alpha'}

\Title{\vbox{\baselineskip12pt
\hbox{AS-ITP-98-11}
\hbox{EFI-98-49}}}{\vbox{\centerline{Large N Strong/Weak Coupling Phase
Transition }
\vskip12pt
\centerline{and the Correspondence Principle}
}}
\vskip16pt
\centerline{Yi-hong Gao$^1$ {\it and} \quad Miao Li$^{1,2}$}
\bigskip
\centerline{\it $^1$Institute of Theoretical Physics}
\centerline{\it Academia Sinica }
\centerline{\it Beijing, China}
\bigskip
\centerline{\it and}
\bigskip
\centerline{\it $^2$Enrico Fermi Institute}
\centerline{\it University of Chicago}
\centerline{\it 5640 Ellis Avenue, Chicago, IL 60637, USA}
\centerline{\it mli@theory.uchicago.edu}

\vskip1.2cm
We argue that the large N strong/weak phase transition is a generic
phenomenon in a finite temperature supersymmetric Yang-Mills theory
of maximal supersymmetry. ${\cal N}=4$, $D=4$ SYM is the canonical
example, where we also argue that the large N Hawking-Page phase
transition disappears for a sufficiently small coupling. 
 The Hawking-Page transition
temperature is lowered by the first $\ap$ correction. Physically,
the strong/weak phase transition is identified with the correspondence
point of Horowitz and Polchinski. We also try to construct toy
models to demonstrate the large N phase transitions, with limited
success.

\Date{Oct. 1998}

\nref\bfss{T. Banks, W. Fischler, S. Shenker and L. Susskind,
hep-th/9810043.}
\nref\jm{J. M. Maldacena, hep-th/9711200.}
\nref\gkpw{S. S. Gubser, I. R. Klebanov and A. M. Polyakov, hep-th/9802109;
E. Witten, hep-th/9802150.}
\nref\miao{M. Li, hep-th/9807196.}
\nref\ew{E. Witten, hep-th/9803131.}
\nref\hpol{G. T. Horowitz and J. Polchinski, hep-th/9612146,
Phys. Rev. D55 (1997) 6189.}
\nref\bg{T. Banks and M. Green, hep-th/9804170, J. High Energy Phys.
05 (1998) 002.}
\nref\itzh{N. Itzhaki, J. M. Maldacena, J. Sonnenschein and S.
Yankielowicz, hep-th/9802042, Phys. Rev. D58 (1998) 046004.}
\nref\gross{D. J. Gross, talk at Strings' 98, http://www.itp.ucsb/online
/strings98/gross.} 
\nref\dvv{R. Dijkgraaf, E. Verlinde and H. Verlinde, hep-th/9703030.}
\nref\lms{M. Li, E. Martinec and V. Sahakian, hep-th/9809061.}
\nref\bkr{J. L. F. Barbon, I. I. Kogan and E. Rabinovici,
hep-th/9809033.}
\nref\hp{S. Hawking and D. Page, Comm. Math. Phys. 87 (1983) 577.}
\nref\gkp{S. S. Gubser, I. R. Klebanov and A. W. Peet,
hep-th/9602135.}
\nref\largen{J. Koplik, A. Neveu and S. Nussinov, Nucl. Phys. B123
(1977) 109; W. T. Tutte, Can. J. Math. 14 (1962) 21.}
\nref\thoof{G. 't Hooft, in ``Progress in gauge field theory'', eds.
G. 't Hooft et al. Plenum Press (1984).}
\nref\gkt{S. S. Gubser, I. R. Klebanov and A. A. Tseytlin, 
hep-th/9805156.}
\nref\pt{J. Pawelczyk and S. Theisen, hep-th/9808126.}
\nref\dg{D. Gross, hep-th/9212149, Nucl. Phys. B400 (1993) 161.}
\nref\ms{J. Maldacena and A. Strominger, hep-th/9804085.}
\nref\gw{D. Gross and E. Witten, Phys. Rev. D21 (1980) 446.}
\nref\dk{M. R. Douglas and V. A. Kazakov, hep-th/9305047, Phys. Lett.
B319 (1993) 219.}
\nref\gv{F. Gliozzi and M. A. Virasoro, Nucl. Phys. B164 (1980) 141.}

\bigskip

\newsec{Introduction}

Supersymmetric Yang-Mills theories (SYM) exhibit rich dynamics.
The study of matrix theory \bfss\ and Maldacena's conjecture \refs{\jm,
\gkpw} reveals much of the unusual physics of SYM with maximal
supersymmetry in the large N limit. Small energy gaps, different
energy scales are only a few examples of many intriguing properties
of these theories. While it is surely a great hint that both matrix
theory and AdS/CFT correspondence encode gravitational physics
in the large N limit, the major technical obstacle to our 
understanding of these models is to truly resolve the large N problem.

Another surprise the AdS/CFT correspondence brings to us is the 
possibility of solving the confinement problem of QCD in the large
N limit. The hope might be hindered by a large N strong/weak
phase transition \miao. According to \ew, D4-brane and D3-brane
theories when compactified on a Euclidean circle lead to QCD
and 3D pure Yang-Mills theory, provided the radius of this circle
is much smaller than the intrinsic scale of the lower dimensional
theory. The latter can be viewed as the low energy sector of the original
theory at a finite temperature. The strong/weak phase transition
of the free energy reflects the same physics of this low energy
sector in the $D=4$ case, since its occurrence is independent of the 
temperature. One can not completely ignore the underlying 4 dimensional
theory at the transition point, since the 3D YM scale $\lambda_3=
\lambda T\sim T$, the KK modes have the same energy scale as the
3D QCD string tension.
While in the $D=5$ case, although the phase transition
is in terms of temperature, so long when the M-circle scale
is small enough, the transition temperature can be very
high, again the phase transition will be reflected by physics of
the low energy sector, namely QCD.
The existence of the phase transition implies that to 
study the weak coupling regime, one must study the weak coupling regime
directly. The weak coupling physics can not be obtained by the
analytic continuation of the strong coupling physics.

The study of \miao\ was conducted in the infinite volume limit,
and is limited to $D=4$ case. We shall generalize it to the
finite volume case, and point out that the strong/weak phase
transition also exists on $S^3$. The phenomenon is also a general
feature of SYM of maximal supersymmetry in various dimensions.
The phase transition point is identified with the correspondence
point of \hpol. This makes the definition of the correspondence
point precise. For the $p+1$ dimensional SYM, if $p\ne 3$,
we predict the phase transition temperature $T_c\sim \lambda^{1/(3-p)}$.
Our study of phase transition is rather qualitative.
To understand dynamic details, one need to study both the weak
coupling side and the strong coupling side to demanding precision.
There is no tool available to enable us to do so.

We shall point out the connection between the strong/weak phase
transition and the correspondence principle in the next section.
In the subsequent section, we argue that the large N first order
Hawking-Page phase transition disappears in the weak-coupling
regime ($D=4$), thus there is a phase transition of phase transition,
which is just the strong/weak phase transition. The correction to
the Hawking-Page transition temperature due to the first $\ap$
correction is found in sect.4, where we determine the critical
coupling $\lambda_c$ using this result. In the final section we
design some toy models to explain large N phase transitions.
We show it is quite hard to come up with a good model to demonstrate
a first order phase transition. We point out possible missing
physics in these models.

\newsec{Strong/weak phase transition and correspondence principle}

It was argued in \miao\ that the free energy of ${\cal N}=4$, $D=4$
SYM at a finite temperature is not a smooth function of the coupling
constant $\lambda =g^2_{YM}N$. To determine the nature of this
phase transition, such as its order, is beyond the power of the
currently available tools.

Denote the critical value of $\lambda$ by $\lambda_c$. For a 
$\lambda <\lambda_c$, one can trust the SYM perturbative expansion.
In this regime, one does not expect the origin $\lambda =0$ be
an essential singularity, given the fact that the underlying 
zero-temperature theory is finite, and the instanton effects are
suppressed in the large N limit \bg. On the other hand, for 
a $\lambda>\lambda_c$, one need to use the whole string theory
on manifold $AdS_5\times S^5$. This reminds us the correspondence
principle of Horowitz and Polchinski \hpol. They postulate that
there exists a correspondence point at which the maximal curvature
invariant at the horizon becomes of the string scale. For a smaller
curvature, the system is in its black hole phase, and the 
semi-classical geometry is reliable in string theory. One can
study the closed string theory in this background.
For a larger curvature, the geometry loses its meaning in string theory,
now the D-brane perturbation theory takes over. Our specification
of a phase transition defines precisely what the correspondence
point is. 

Here we must note that the following argument is not
meant to prove the existence of a phase transition, rather, we
intend to argue that if there exists a phase transition, it is natural
to identify it with the correspondence point. This identification
has proven to be useful in recent work \refs{\lms, \bkr}.

The connection between the strong/weak phase transition and the
correspondence principle looks puzzling at the first sight.
For a standard Schwarzschild black hole, the curvature invariants
would be proportional to $1/r_0^2$, where $r_0$ is the horizon size.
This quantity depends on the Hawking temperature. We know that
in a conformally invariant theory with an infinite volume, there
can be no phase transition in terms of temperature.
The fact that we are applying the correspondence principle to a
AdS black hole saves the day. We will show, that all the curvature
invariants in such a background are independent of $r_0$, so
long as $r_0>0$. Thus the maximal curvature is determined by the
AdS size, $R=(2\lambda)^{1/4}l_s$. Now $R\sim l_s$ if $\lambda
\sim O(1)$. This is consistent with the fact that $\lambda_c
\sim O(1)$.

The metric of the $AdS_5$ black hole, with a boundary $R^3\times S^1$, is \jm
\eqn\adsm{ds^2= {r^2\over R^2}[(1-{r_0^4\over r^4})
dt^2+\sum_{i=1}^3dx_i^2] +{R^2\over r^2}(1-{r_0^4\over r^4})^{-1}dr^2.}
With this metric, it is not straightforward to see that all the
curvature invariants are independent of $r_0$. To do so, we follow
\miao, to rewrite the metric in a different coordinates system.
Introduce
\eqn\ctrans{r^4=r_0^4(1+b^2), \quad t={R^2\over 2r_0}\psi ,}
the metric becomes
\eqn\mtrans{ds^2={R^2\over 4}\left( {db^2\over 1+b^2}+
{b^2d\psi^2\over\sqrt{1+b^2}}+\sqrt{1+b^2}\sum_{i=1}^3dx_i^2
\right),}
where we also rescaled $x_i$. The metric in the parentheses is
universal, thus its curvature invariants are universal too. We
conclude that the curvature invariants of the AdS black holes
are independent of $r_0$, and determined only by the AdS size
$R$.

The universal metric \mtrans\ is applicable only when $r_0>0$.
The AdS black hole solution is certainly different from that of
AdS. The coordinates transformation \ctrans\ is singular in the
$r_0=0$ limit. In other words, the metric \mtrans\ describes
the very near-horizon region in the $r_0\rightarrow 0$ limit.

The observation on the connection between the large N strong/weak
phase transition and the correspondence point generalizes to
other SYM with the maximal supersymmetry in various dimensions.
For $D\ne 4$, the YM coupling constant is dimensionful, and
the construction of a dimensionless constant involves both
the coupling constant and the energy density. Thus the strong/weak
phase transition becomes of a finite density or a finite temperature
phase transition. Let $\lambda =g^2_{YM}Nd_p$, where $d_p$ is a 
numeric factor depending only on $p$. 
The near-horizon metric of a nonextremal Dp-brane is \itzh
\eqn\nonex{ds^2=\alpha'\left( f dt^2+{U^{(7-p)/2}\over \sqrt{\lambda}}
\sum dx_i^2+f^{-1}dU^2
+\sqrt{\lambda}U^{(p-3)/2}d\Omega_{8-p}^2\right),}
where
\eqn\appf{ f={U^{(7-p)/2}\over \sqrt{\lambda}}\left(1-{U_0^{7-p}
\over U^{7-p}}\right).}
The dilaton field is given by
\eqn\dila{e^\phi =(2\pi)^{2-p}g^2_{YM}\left( {\lambda\over U^{7-p}}
\right)^{(3-p)/4}.}
For $p<3$, the string coupling becomes stronger when $U_0$
is smaller. For $p>3$, it becomes weaker when $U_0$ is smaller.

As in the $p=3$ case, use
\eqn\nmtr{U^{7-p}=U^{7-p}_0(1+b^2),\quad
t={2\over 7-p}U^{-1}_0\sqrt{\lambda_{eff}}\psi,}
where $\lambda_{eff}=\lambda U_0^{p-3}$, the metric is put into
\eqn\ntr{\eqalign{ds^2=&{4\ap\sqrt{\lambda_{eff}}  \over (7-p)^2}
\left((1+b^2)^{{3p-17\over 14-2p}}db^2+{b^2d\psi^2\over\sqrt{1+b^2}}
+{(7-p)^2\over 4}(1+b^2)^{{p-3\over 14-2p}}d\Omega_{8-p}^2\right) \cr
&+{U_0^{(7-p)/2}\over \sqrt{\lambda}}(1+b^2)^{(7-p)/2}\sum dx_i^2.}}
We see that the curvature is controlled by $\lambda_{eff}$.
When this quantity becomes order 1, the maximal curvature
becomes of the string scale. Since the Hawking temperature is
\eqn\htem{T=(\lambda_{eff})^{-1/2}U_0({7-p\over 4\pi}),}
the temperature can be identified with $U_0$ at the correspondence point. 
This in turn implies
that the effective coupling at the correspondence point
is $\lambda_{eff}=\lambda T^{p-3}\sim 1$, and the
phase transition occurs at
\eqn\ctem{T_c\sim \lambda^{{1\over 3-p}}.}

The dilation field at the horizon is
\eqn\dilat{e^\phi\sim {1\over N}(\lambda_{eff})^{{7-p\over 4}}.}
it is of order $1/N$ at the phase transition point. The large
N approximation becomes quite good, and on the string side one
is studying free strings propagating in the black hole background.

When the above analysis applied to $p=4$ case, we predict that
the phase transition occurs at $T_c\sim \lambda^{-1}$.  For higher
temperatures, $\lambda_{eff}$ is larger, and we are in
the supergravity phase. For lower temperatures, we need to apply
the perturbation theory of SYM or the underlying M5-brane theory.
To study QCD following \ew, temperature $T$ must be adjusted over
the induced QCD scale $T\gg \Lambda_{QCD}$. At the QCD scale, the
four dimensional QCD coupling $\lambda_4=\lambda T$ receives
quantum corrections, but it is still of order 1. Thus, the situation
is close to the transition point. We believe that it is located 
at the small coupling side. One of hints is the observation that
the Regge trajectory in the strong coupling regime displays 
very different pattern as expected of QCD \gross.

The 5 dimensional Yang-Mills coupling is given by $g_{YM}^2=R_{11}$,
where $R_{11}$ is the radius of the M-circle around which the M5-branes
are wrapped to obtain D4-branes.  Thus the phase transition occurs
at $\beta/R_{11}\sim N$.

Another interesting case is $p=1$. The transition temperature is
given by $T_c\sim \lambda^{1/2}$. For $T\gg T_c$, the theory is
strongly coupled. If we switch the role of time and space, the
theory becomes the 2D SYM living on a circle of circumference
$\beta$. This can be related to matrix string theory \dvv. In 
the strong coupling regime, the theory is effectively a conformal
orbifold model, and the physics is dominated by long-strings.
For $1/T$ small enough, one can ignore fermions, since
fermions are anti-periodic. The theory is effectively bosonic.
When $T$ is lower than $T_c$, the theory is
nonabelian. The phase transition separates the abelian orbifold
phase of the bosonic sector and the nonabelian phase  of the
NS sector.

We turn to $AdS_5$ with boundary $S^3\times S^1$ in the next two sections.
For discussions on the phase diagrams with other topologies, see
\lms\ and \bkr.

\newsec{Disappearance of Hawking-Page phase transition}

Hawking and Page found some time ago \hp, that there are two manifolds
for a given temperature on AdS space, provided the temperature is
not too low. This was re-interpreted by Witten as a large N phase
transition in the ${\cal N}=4$ $D=4$ SYM \ew. This is a finite volume
effect, and occurs only when the SYM lives on $S^3\times S^1$ or
$S^2\times R\times S^1$.

The AdS black hole metric with boundary $S^3\times S^1$ is
\eqn\adsn{ds^2=(1+{r^2\over R^2}-{r_0^2\over r^2})dt^2
+(1+{r^2\over R^2}-{r_0^2\over r^2})^{-1}dr^2+r^2d\Omega_3^2.}
For a given Hawking temperature, there are two black holes, one
small, one large. The smaller one can not be regarded as a
thermal state in the boundary Yang-Mills theory, since its
entropy is not an extensive quantity. The horizon radii of these
two black holes are
\eqn\radii{r_{\pm}={\pi R^2T\over 2}\left(1\pm (1-{2\over (\pi TR)^2}
)^{1/2}\right).}

Use the standard formula for the Bekenstein-Hawking entropy
$A/4G_5$, the entropy reads
\eqn\entr{S_\pm={\pi^2N^2\over 2}(2\pi^2R^3)T^3\left({1\over 2}\pm{1\over 2}
(1-{2\over (\pi TR)^2})^{1/2}\right)^3,}
where we used formulas $G_5=G_{10}/\pi^3 R^5$, and 
$G_{10}=2^3\pi^6g_s^2(\ap)^4$. In the large volume or high temperature
limit, $S_-$ is not an extensive quantity. 
Compare $S_+$ with the one
in the infinite volume limit \gkp, we find that $V_3=2\pi^2 R^3$,
namely the radius of $S^3$ is also $R$. Note that the precise
definition of the radius of $S^3$ depends on the choice of time, 
or that only $TR$ is convention independent.

There is a finite size correction to the entropy formula. And formula
\entr\ displays a branch cut at $\pi TR=\sqrt{2}$. There is a minimal
Hawking temperature. However, the Hawking-Page transition temperature
is $\pi T_cR=3/2$. This value is obtained by demanding the subtracted
action vanishing, thus $r_+^2=R^2$.
The critical temperature is higher than the minimal temperature, thus
there is no particular physics associated with the minimal temperature.

The free energy can be obtained by using the relation $S=\beta^2
\p_\beta F$. 

One can trust the above picture only when classical supergravity is
valid, thus the radius of $S^3$ is much larger than the string
scale. In terms of SYM, this requires $\lambda \gg 1$. We now argue
that for a small $\lambda$, the first order phase transition
disappears. Denote the phase transition temperature by $T_c(\lambda )$,
it is a function of $\lambda$. In the large $\lambda$ limit,
$T_c R$ tends to a constant, $3/(2\pi)$. If the phase transition 
disappears for small $\lambda$, there must be a critical $\lambda_c$,
at which $T_c=0$. This phase transition of phase transition must
be the strong/weak phase transition.

Apparently, $T_c=0$ at $\lambda =0$. The theory is free, and the free
energy can be computed by calculating determinants of free fields.
The number of free fields is of order $O(N^2)$, so the free energy
scales as $F\sim N^2$, and we always have the high temperature
phase. For a small $\lambda$, to include effects of interaction,
one employs the loop expansion in $\lambda$, and in the large N limit,
it is sufficient to count the planar diagrams. At each loop order
$\lambda^n$,
the number of planar diagrams grows as $c^n$ \refs{\largen, 
\thoof}. There is no UV divergence,
since the underlying zero temperature theory is finite. There is
also no IR divergence, since both the space and the time have a finite
size. Thus, the systematic loop expansion of the free energy assumes
the following form
\eqn\freel{F=-N^2T^4R^3\sum_{n=0} \lambda^nf_n(TR),}
where $f_n$ depends only on the dimensionless combination $TR$. 
We do not expect effects such as UV or IR renormalons, the above series
ought to be convergent for a small enough $\lambda$, and the series
typically has a finite convergent radius.

By a similar argument, each perturbative series at a given genus
also has a finite convergent radius. Now it is possible that a
singularity is located on the positive axis of $\lambda$, denote
this by $\lambda_c$. Beyond this point, one can no longer trust
the gauge theory perturbative expansion, and one must invoke other
techniques such as the closed string theory on the AdS space to
compute the free energy. for $\lambda<\lambda_c$, the free energy
{\it always} scales as $N^2$, and there is no Hawking-Page phase
transition. The critical $\lambda_c $ is determined by the
asymptotic behavior of $f_n(TR)$. It is still possible that
$\lambda_c$ is a function of $TR$. Consider the limit
$TR\rightarrow \infty$. In such a limit, as argued in \miao,
there is a finite critical $\lambda_c$. When one lowers $TR$,
it is possible that $\lambda_c$ is also lowered, but cannot be
pushed all the way to zero. For instance, assume that the convergent radius
is proportional to  $T^\alpha$ with a positive $\alpha$, then the 
asymptotic form of $f_n\sim (1/TR)^{n\alpha}$. Analyzing the Feynman
diagrams, such a factor can come only from the propagators of
the zero modes. Now on $S^3$, there is no zero mode at all.
Put in another way, there is a mass gap in the SYM on $S^3$ which
is proportional to the smallest scaling dimension.
Still denote the minimum of $\lambda_c(TR)$
by $\lambda_c$. We conclude that there is no large N
Hawking-Page phase transition below $\lambda_c$.

The existence of Hawking-Page phase transition indicates that
there exists a finite $\lambda_c$, above which the perturbative SYM
ceases to be valid, and the AdS picture takes over. We shall
show in the next section that the Hawking-Page temperature is
lowered by the first $\ap$ correction.
 
\newsec{$\ap$ correction to the thermodynamic quantities}

The correction to thermodynamics by the first $\ap$ correction
was done in \gkt, in the infinite volume limit, where it was found that 
the correction to the entropy is positive. There are two ways
to compute this correction. The simpler way is to substitute the
Riemann curvature of the metric \adsn\ directly to the term
$(\ap)^3 R^4$ to compute the subtracted action. Another is to take
the corrected metric, dilaton into account. The two calculations
give the same result. The full answer of the corrected metric was
found in \pt, again in the infinite volume limit.

We will adopt the simpler method, hoping that the corrected metric
will not change the final results. The full form of the $(\ap)^3R^4$
is 
\eqn\cact{\Delta I=-{1\over 16\pi G_{10}}\int {1\over 8}\zeta (3)
(\ap)^3 W+\dots,}
where the dots denote terms depending on other fields, and
\eqn\wform{W=C^{hmnk}C_{pmnq}C_h^{rsp}C^q_{rsk}+{1\over 2}
C^{hkmn}C_{pqmn}C_h^{rsp}C^q_{rsk}.}

Introducing angular variables $\chi, \theta, \phi$ on $S^3$,
after some calculations, we obtain those nonvanishing  components of
the Weyl tensor
\eqn\weyl{\eqalign{C^{ab}_{\quad cd}&=3X\epsilon^{ab}\epsilon_{cd},
\qquad a,b=t,r,\cr
C^{ai}_{\quad bj}&=-X\delta^a_b\delta^i_j, \qquad 
i,j=\chi, \theta, \phi,\cr
C^{ij}_{\quad kl}&=X(\delta^i_k\delta^j_l-\delta^j_k\delta^i_l),}}
where $X=r_0^2/r^4$. The difference between our result and that
of \gkt\ is that $X$ is independent of $R^2$ in our case.

It is straightforward to substitute the above result into \cact, the
result is
\eqn\ccact{\Delta I=-{15\pi\zeta (3)\over 64}{(\ap)^3r_0^8\beta
\over G_5 r_+^{12}}.}
Using the formula $\Delta S=\beta^2\p_\beta \Delta F$ we obtain
\eqn\centr{\Delta S={15\pi^2\zeta (3)\over 32}N^2(2\lambda)^{-3/2}
V_3T^3[{3\over 2}-{1\over 2}(1-{2\over (\pi TR)^2})^{1/2}]^3
[3-(1-{2\over (\pi TR)^2})^{-1/2}].}
In the large volume limit we recover the result of \gkt. Note that
the last factor approaches minus infinity when $T$ approaches its
minimal value. At the uncorrected transition temperature $T_c$, it becomes
zero.

Demanding the subtracted action be zero, we calculate the corrected
Hawking-Page temperature, up to the leading order $(\ap)^3$
\eqn\ctem{\pi RT_c={3\over 2}-{15\over 2}\zeta (3)(\ap)^3/R^6
={3\over 2}-{15\over 2}\zeta (3)(2\lambda )^{-3/2}.}
As expected, the corrected temperature is lower. The correction
to the entropy \centr\ at this point is weighted by $\lambda^{-3}$.
The last factor will never be too negative to render the total
entropy become negative.

Although the critical temperature is lowered, we can not conclude that
$T_c$ will drop to zero at a finite $\lambda_c$. If we are optimistic
and assume the higher $\ap$ terms can be ignored at $\lambda_c$,
Eq.\ctem\ can be used to determine $\lambda_c$:
\eqn\lamc{\lambda_c= {1\over 2}(5\zeta (3))^{2/3}+\dots .}
This value is greater than 1.

As in sect.2, the connection between the strong/weak phase transition
point and the correspondence point is still valid. For the larger
black hole, since $r_+\ge R/\sqrt{2}$, the maximal curvature is
always controlled by $R$. The smaller black hole is more like
a Schwarzschild black hole. Its specific heat is negative, and can
be described at best as a meta-stable coherent state in the SYM theory. 
The horizon can be arbitrarily small, and the maximal curvature
at the horizon, unlike the case discussed in sect.2, can be arbitrarily
large.

\newsec{Some toy models}

We design some artificial toy models in this section to explain
the occurrence of the large N phase transitions. To our knowledge,
the exactly solvable models such as those studied in \gw\ and
\dk\ exhibit different, although similar, phase transitions to
that of Hawking and Page. Those models are all effectively 
one-matrix models, and the phase transition is of the third order. The
Hawking-Page transition is a first order one, since the free energy
scales differently in $N$ in the low and high temperature phases.

Consider a conformal field theory defined on spacetime $S^d\times R$.
Let $H$ denote the Hamiltonian on $S^d$. The free energy is
given by
\eqn\freed{F=-\ln \left(\sum e^{-\beta H}\right).}
The metric on $S^d\times R$ reads
\eqn\metric{ds^2=d\tau^2 +R^2d\Omega_d^2,}
where $R$ is the radius of $S^d$. This metric is conformal to a flat
metric on $R^{d+1}$. Use the new coordinate $r=exp(\tau/R)$, the
metric is mapped to
\eqn\nmetric{ds^2={R^2\over r^2}\left( dr^2+r^2d\Omega_d^2\right),}
the new metric in the parentheses is the flat one on $R^{d+1}$.
Now the time translation $\tau\rightarrow \tau + a$ gets translated
into rescaling of $r$: $r\rightarrow \exp (a/R) r$. Thus the Hamiltonian
on $S^d$ is identified with $D/R$, where $D$ is the dilatation operator
on $R^{d+1}$ acting at the origin. For simplicity, we still use
$\beta$ to denote the combination $\beta/R$. The task of computing
the free energy now becomes of counting scaling dimensions:
\eqn\count{F=-\ln\left(\sum e^{-\beta D}\right).}

A large N quantum field theory is very complicated, so we will construct
simplest possible large N models, one matrix models or 
multiple-matrix models. Given a Hermitian
one matrix $\phi$, there are only $N$ independent single trace operators
$\tr\phi^n$, $n=1,\dots N$. A ``multi-particle'' state, similar
to those considered in \ms, is given by the  product of a string of such
single trace operators, 
\eqn\strin{\tr\phi^{n_1}\dots \tr\phi^{n_l}.}
Assign a scaling dimension $\Delta (n,\lambda)$ to operator 
$\tr\phi^n$, where we introduced a coupling constant $\lambda$.
The scaling dimension of a multi-particle operator is generally
complicated. To simplify the situation, we assume the scaling
dimension of the general operator be given by
$\sum \Delta (n_i,\lambda)$. The partition function is
then 
$$\prod_{n=1}^N(1-q^{\Delta (n,\lambda)})^{-1},$$
where $q=\exp (-\beta)$. The free energy
\eqn\frees{F_N=\sum_{n=1}^N\ln\left(1-q^{\Delta (n,\lambda)}\right).}

The first model is specified by $\Delta (n,\lambda)=n$. All operators
mimic the chiral primary operators. In such a case, we expect no
phase transition when the temperature is varied. This model looks
the same as the 2D QCD string on a torus \dg. The free energy is
calculated by expanding the logarithmic 
\eqn\nfree{F_N=-\sum_{k=1}^\infty{q^k\over k}(1-q^k)^{-1}
+\sum_{k=1}^\infty{q^k\over k}(1-q^k)^{-1}q^{Nk}.}
There is no $1/N$ corrections. The second sum is due to the instanton
contribution. Each term can be interpreted as the k-instanton
contribution.

Our second model is given by $\Delta (n,\lambda)=a(\lambda)\ln n
+b(\lambda)$ with $b>0$. If $a(\lambda)\ne 0$, for large $n$, the scaling
dimension receives large quantum correction. These operators
mimic operators corresponding to stringy states. $b(\lambda)>0$
is to ensure the positivity of the first scaling dimension.

Again, expanding the logarithmic in \frees, we have
\eqn\rzeta{F_N=-\sum_{k=1}^\infty{e^{-k\beta b}\over k}
\sum_{n=1}^N n^{-k\beta a}.}
Apparently, the behavior of the finite sum depends crucially
on the value of $\beta a$. The dominant contribution to $F_N$
comes from $k=1$.
\eqn\divi{\eqalign{\sum_{n=1}^N n^{-\beta a}=\matrix{
N^{1-\beta a}, &\beta a\ne 1\cr
\ln N, &\beta a=1}}}

The model exhibits kind of phase transition. When $\beta a<1$,
or $T> a$, the free energy scales as a positive power of $N$;
it tends to a constant when $T<a$, and grows as $\ln N$ at
$T=a$. So $T_c=a$ is the critical temperature. Unfortunately,
there is no standard large N expansion. To see the nonanalyticity
in $N$, we write
\eqn\exac{F_N=-\sum_{k=1}^\infty {e^{-k\beta b}\over k}
[\zeta (k\beta a)-\zeta(k\beta a,N+1)]=f(1)-f(N+1).}
the first term is independent of $N$, and is regular provided
$\beta a\ne 1/k$. To examine the second term, use the Hermite
formula
\eqn\herm{\eqalign{\zeta(k\beta a,N+1)
=&{1\over 2} (N+1)^{-k\beta a}+{1\over k\beta a-1}(N+1)^{1-k\beta a}
\cr
&+2\int_0^\infty dy ((N+1)^2+y^2)^{-k\beta a/2}\sin(k\beta a\theta)
(e^{2\pi y}-1)^{-1},}}
where $\theta =\arctan(y/(N+1))$. One finds
\eqn\exact{\eqalign{f(N+1)=&f_{crit}(N+1)+(N+1/2)\ln [1-
(N+1)^{-\beta a}e^{-\beta b}]\cr
&-\int_0^\infty dy (e^{2\pi y}-1)^{-1}\ln\{ [1-(\sqrt{(N+1)^2+y^2}
e^{-ia\theta+b})^{-\beta}]
\cr
&[1-(\sqrt{(N+1)^2+y^2}
e^{ia\theta+b})^{-\beta}]^{-1}\},}}
where the first term and the second term control the large N behavior.
The first term is given by
\eqn\non{f_{crit}(N+1)=\sum_{k=1}^\infty {\beta a \over k\beta a-1}
e^{-k\beta b}(N+1)^{1-k\beta a}.}

We are convinced that it is impossible to construct a one-matrix model
to exhibit both a first order phase transition and the usual large
N expansion. This leads us to examine multiple-matrix models.
As soon as there are more than one matrix, the number of single-particle
states at a given level starts to proliferate. For instance, consider
two matrices. A general single-particle state is
\eqn\sing{\tr \left(\phi_1^{m_1}\phi_2^{n_1}\dots \phi_1^{m_l}\phi_2^{n_l}
\right).}
Define the level $n=\sum (m_i+n_i)$. At a given level, the number of
the single-particle states grows at least as the partition number
$P(n)$, since there is no longer permutation symmetry among $m_i$ and
$n_i$, only the cyclic symmetry is retained. This growth of the
single-particle states is the generic property of a string theory.

For finite N, one must implement constraints on multiple-particle
states. These constraints are known as the Mandelstam constraints,
and as the stringy exclusion principle in the modern context \ms.
It is difficult to implement these constraints efficiently. We will
introduce a cut-off in the same fashion as in \frees. The class of
our multiple-matrix models is specified by
\eqn\mult{Z_N=\prod_{n=1}^N(1-q^{\Delta (n,\lambda)})^{-p(n)},}
where we assumed that all the single-particle states at a given level
$n$ have the same scaling dimension $\Delta (n,\lambda )$, 
and the multiplicity is given by $p(n)$. Needless to say, these models 
are over-simplified multiple-matrix models.

We show that some physics is missing in these toy models, so that
it is impossible to come up with a first order phase transition.
First, we take $\Delta (n, \lambda)=a(\lambda)\sqrt{n}+b(\lambda)$.
The factor $\sqrt{n}$ is introduced according to \gkpw. As we argued
before, the growth of $p(n)$ is at least close to the partition
number $P(n)$, thus $p(n)\sim \exp (c\sqrt{n})$. This introduces
a problem, namely there exists a limiting temperature, the Hagedorn
temperature. Above the Hagedorn temperature, the free energy will
increase as an exponential of $\sqrt{N}$. This is not allowed. So
we shall assume that due to Mandelstam constraints, the growth of
$p(n)$ is weaker than $\exp (c\sqrt{n})$.

Take the logarithmic of \mult, the sum can be approximated by an
integral, in the large N limit. The upper bound of the integral
can be pushed to infinity, since we assumed that there is no
limiting temperature. Thus, the free energy is given by, in the
large N limit
\eqn\intf{F=\int dx p(x, N)\ln\left(1-e^{-\beta a \sqrt{x}-\beta b}
\right),}
where we assumed that $p(x,N)$ in general is a function of $N$ too.
If there were a critical $\beta_c$ at which a first order phase
transition happens, the first derivative of $F$ would diverge. 
Take the first derivative
\eqn\fdf{{dF\over d\beta}=\int dx {(a \sqrt{x}+b)p(x,N)\over
e^{\beta a \sqrt{x}+\beta b}-1}.}
We now require that the above integral diverge for a $\beta_c$.
The divergence should not come from $x=\infty$. If it comes
from $x=x_0$, then $x_0$ must be a singularity of $p(x,N)$,
since the other factor is regular everywhere. In this case,
the integral will diverge for all $\beta$, not just for a single
$\beta_c$. 

We conclude that our simple-minded toy models do not capture the
physics of ${\cal N}=4$, $D=4$ SYM at a finite temperature. Many
things might have gone wrong in our models. It is possible
that solution to the Mandelstam constraints can not be mimicked 
by our ansatz. It is also possible that the assumption that the
scaling dimension of a multiple-particle operator is just
$\sum \Delta (n_i\lambda )$ is badly wrong. It remains a challenging
problem to construct a more realistic model to demonstrate
both the first order Hawking-Page phase transition, as well
as the large N strong/weak coupling phase transition.

\noindent{\bf Acknowledgments} We thank H. Liu, A. Rajaraman and
other participants to the M-theory workshop at  Aspen Center of
Physics for discussions. We also thank M. Yu for a useful discussion. 
Some of the work presented here was done while M.L. participated
a workshop at Aspen Center of Physics. Its warm hospitality is gratefully
acknowledged. This work was completed during a visit of M.L. at ITP and 
CCAST at  Beijing, and their warm hospitality is equally gratefully
acknowledged.
The work of M.L. was supported by DOE grant DE-FG02-90ER-40560 and NSF grant
PHY 91-23780.

\listrefs
\end